\documentclass[%
 aip,
 amsmath,amssymb,
 reprint,%
floatfix,
]{revtex4-1}

\usepackage{amsfonts}
\usepackage{amsmath}
\usepackage{amssymb}
\usepackage{graphicx}
\usepackage{mathrsfs}
\usepackage{color}
\usepackage{bbold}
\usepackage{CJK}
\usepackage{times}
\usepackage{hyperref}

\usepackage{graphicx}% Include figure files
\usepackage{dcolumn}% Align table columns on decimal point
\usepackage{bm}% bold math
\usepackage[utf8]{inputenc}
\usepackage[T1]{fontenc}
\usepackage{mathptmx}
\usepackage{color}

\begin{document}

\title[]{Ultra-Highly Linear Magnetic Flux-to-Voltage response in Proximity-based Mesoscopic bi-SQUIDs}

\author{Giorgio De Simoni}
\email{giorgio.desimoni@sns.it}
\affiliation{NEST, Istituto Nanoscienze-CNR and Scuola Normale Superiore, I-56127 Pisa, Italy}
%\altaffiliation{Current address: Some other place, Othert\"own,
%Germany}
\author{Lorenzo Cassola}
\affiliation{NEST, Istituto Nanoscienze-CNR and Scuola Normale Superiore, I-56127 Pisa, Italy}
\affiliation{Department of Physics ”E. Fermi”, Università di Pisa, Largo Pontecorvo 3, I-56127 Pisa, Italy}
\author{Nadia Ligato}
\affiliation{NEST, Istituto Nanoscienze-CNR and Scuola Normale Superiore, I-56127 Pisa, Italy}
%\affiliation{TeCIP Institute, Scuola Superiore Sant’Anna, 56124 Pisa, Italy}
\author{Giuseppe C. Tettamanzi}
\affiliation{Institute of Photonics and Advanced Sensing and School of Physical Sciences,
The University of Adelaide, Adelaide SA 5005, Australia}
\author{Francesco Giazotto}
\email{francesco.giazotto@sns.it}
\affiliation{NEST, Istituto Nanoscienze-CNR and Scuola Normale Superiore, I-56127 Pisa, Italy}

\preprint{AIP/123-QED}

\begin{abstract}
Superconducting double-loop interferometers (bi-SQUIDs) have been introduced to produce magnetic flux sensors specifically designed to exhibit ultra-highly linear voltage response as a function of the magnetic flux. These devices are very important for the quantum sensing and for signal processing of signals oscillating at the radio-frequencies range of the electromagnetic spectrum. Here, we report an Al double-loop bi-SQUIDs based on proximitized mesoscopic Cu Josephson junctions. Such a scheme provides an alternative fabrication approach to conventional tunnel junction-based interferometers, where the junction characteristics and, consequently, the magnetic flux-to-voltage and magnetic flux-to-critical current device response can be largely and easily tailored by the geometry of the metallic weak-links. We discuss the performance of such sensors by showing a full characterization of the device switching current and voltage drop \textit{vs.}  magnetic flux for temperatures of operation ranging from 30 mK to $\sim 1$ K. The figure of merit of the transfer function and of the total harmonic distortion are also discussed. The latter provides an estimate of the linearity of the flux-to-voltage device response, which obtained values as large as 45 dB. Such a result let us foresee a performance already on pair with that achieved in conventional tunnel junction-based bi-SQUIDs arrays composed of hundreds of interferometers.

\end{abstract}

\maketitle

\section{Introduction}
Superconducting quantum interference devices (SQUIDs) provide the reference standard for magnetic flux detectors \cite{Clarke2004,Barone1982,Kleiner2004,Martinez-Perez2017,Granata2016,Giazotto2008a,Fagaly2015} and for the measurement of all those physical quantities that can be transduced from magnetic to electrical properties. In this sense, SQUIDs can be integrated into larger systems and used for signal processing applications~\cite{Kornev2020}. In their simplest implementation, direct-current (DC) SQUIDs comprise a pair of Josephson junctions (JJ) \cite{Josephson1962} closed in a superconducting ring, whose critical current is modulated by the magnetic flux threading the loop with a periodicity equal to the magnetic flux quantum $\phi_0=h/2e$ \cite{Clarke2004,Barone1982,Doll1961,Deaver1961}.  Because DC SQUIDs sensing abilities are directly linked to the basic principles of quantum mechanics, they can approach the quantum limits of sensing already in their conventional implementations~\cite{Clarke2004}. DC SQUIDs are however characterized by a poor performance in terms of linearity and dynamic range: these issues were routinely circumvented by means of a specific optimization which, at the cost of a drastic reduction of the operation bandwidth, is implemented through the introduction of external feed-back loops~\cite{Clarke2004,Kleiner2004,Fagaly2015}. On the other side, high-frequency open-loop SQUID amplifiers, that were demonstrated to be suitable for applications up to the gigahertz range \cite{Prokopenko2003,Prokopenko2001,Muck2001a,Prokopenko1999,Prokopenko1997,Huber2001,Muck2001b,Muck2001,Muck1999,Andre1999,Muck1998}, exhibit severe limitations due to a significant non-linear distortion\cite{Muck2001b}. A mitigation of these restrictions came from the introduction of a class of superconducting interferometers containing a third JJ connected in parallel with the inductance loop of a DC SQUID. In such devices, namely called bi-SQUIDs~\cite{Kornev2020a,Kornev2011,Kornev2014,Kornev2009}, the third JJ,~is used as additional non-linear element operating in parallel to compensate for the non-linear response of the DC-SQUID resulting in a highly linear voltage response \cite{Kornev2020a,Kornev2011,Kornev2014}. Although such promising premises, Nb bi-SQUIDs with shunted Josephson tunnel junctions \cite{Sharafiev2012}, due to the large junction area and inductance, showed non ideal voltage response, with a linearity performance far from the expected one~\cite{Kornev2020}. Instead, Nb bi-SQUIDs based on arrays of different numbers of interferometers, ranging between tens and hundreds of unit cells, have proven to be extremely effective for low-noise signal amplification and magnetic field sensing, exhibiting excellent performance in terms of linearity of the flux-to-voltage response\cite{Kornev2017,Kornev2017a,Kornev2011a}.
\begin{figure}[ht]
  \includegraphics[width=0.9\columnwidth]{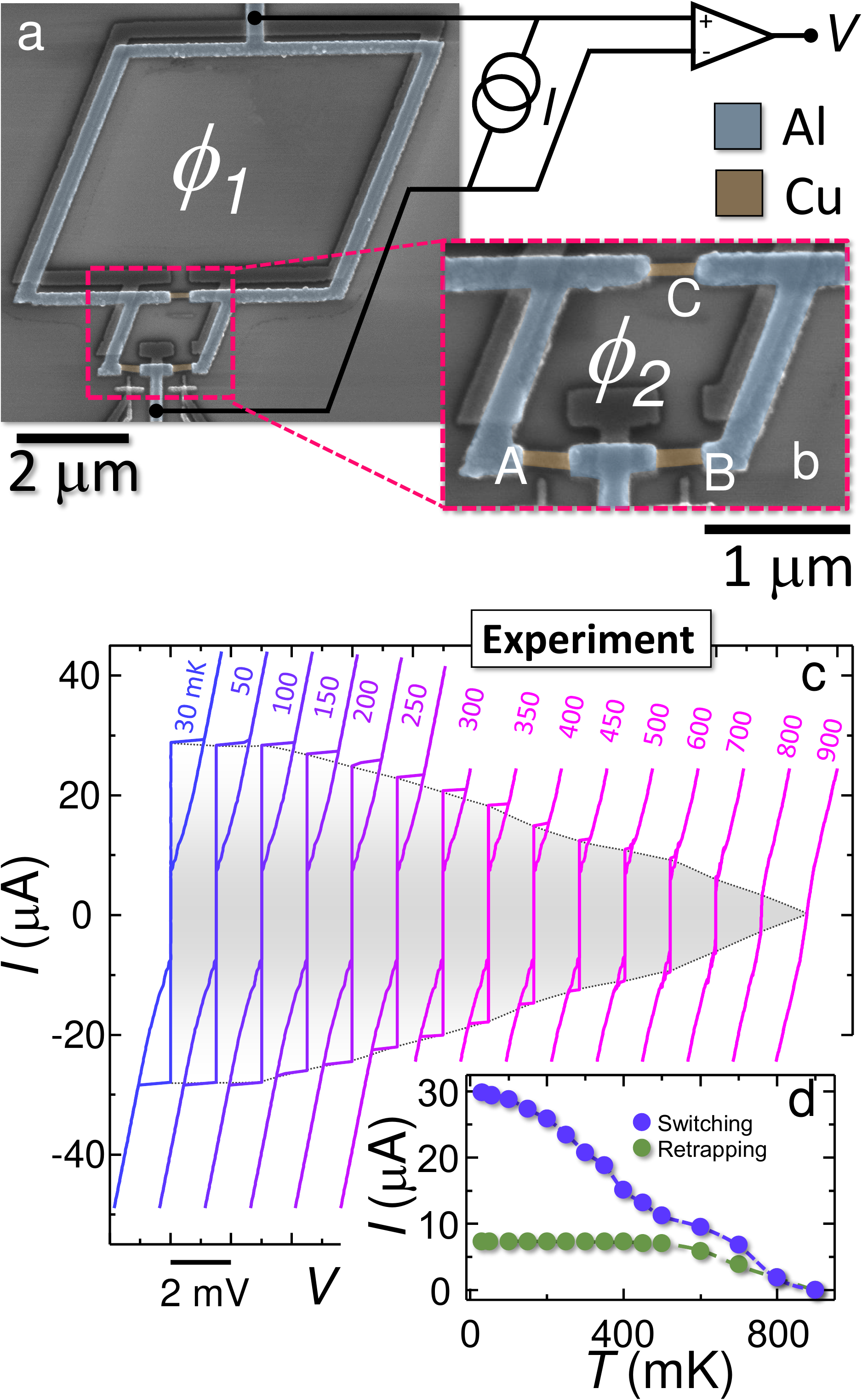}
    \caption{\textbf{Proximity-based gated all-metallic bi- SQUID} \textbf{a}: False-color scanning electron micrograph of a representative double-loop DC superconducting  quantum interference device (bi-SQUID) based on superconductor/normal metal/superconductor (SNS)  proximity Josephson junctions. The Al interferometer ring is coloured in blue. The Cu Josephson weak-links are colored in brown. The 4-wire electrical setup is also shown. \textbf{b}: Higher magnification false-color scanning electron micrograph of the second loop of the bi-SQUID. The Cu weak-links are labelled as A, B, and C. \textbf{c}: Current ($I$) \textit{vs.} voltage ($V$) forward and backward characteristics of a representative bi-SQUID at selected temperatures between 30 mK and 900 mK and at zero magnetic flux. Curves are horizontally offset for clarity. The $I-V$ non-dissipative region is gray-shaded. \textbf{d}: Switching (violet) and retrapping (green) current of the same device of panel a \textit{versus} temperature $T$.}
  \label{fig:fig1}
\end{figure}

Although the great majority of superconducting interferometers used in commercial applications exploit superconductor/insulator/superconductor (SIS) JJs, the DC Josephson effect can be observed in a broad variety of systems~\cite{Likharev1979},
such as in Dayem bridges\cite{Vijay2010,Levenson-Falk2013}, or in weak-links based on a semiconductor~\cite{Giazotto2004,Carillo2006,Giazotto2011a} or a normal metal~\cite{Savin2004a} sandwiched between a pair of superconducting leads (SNS JJs). SNS junctions support a non dissipative current thanks to the proximity effect \cite{Pannetier2000a} stemming from the building of Andreev bound states in the N region~\cite{Belzig1999,McMillan1968}, that lends the superconducting correlations to the electron gas in clean electric contact with the S leads. SNS JJs gained a growing interest in device physics thanks to a negligible parasitic capacitance and, even more, to a convenient and reproducible fabrication process, which allows their critical current  and the functional form of their current-phase-relation\cite{Heikkila2002a,Baselmans1999} to be tailored to specific application needs just through the geometry of the weak link. Similarly to what conventional done for tunnel junctions devices~\cite{Kornev2020}, in bi-SQUIDs based on SNS weak-links, the linearity of the magnetic-field-to-voltage response can be controlled through the ratios of the critical current of the three JJs. The latter, indeed, can be regulated during the fabrication by properly setting the section and the length of the weak-links, and the relative thickness and the size of the overlapping area of the S and N regions. Furthermore, the critical current of SNS JJs can be tuned down to zero by the application of a gate control voltage\cite{DeSimoni2019,DeSimoni2021}. This last feature, which was not demonstrated so far for tunnel junctions, might be exploitable to vary and improve the device performances during its operation~\cite{Tettamanzi2021}. 
 In this work, we demonstrate the operation of  Al bi-SQUIDs  in which  three nano-junctions are implemented through  proximitized mesoscopic Cu weak-links. Although its relatively low critical temperature, Al was chosen due to its efficient proximization capability over copper \cite{DeSimoni2019,DeSimoni2021}, but we emphasize  that this scheme can be easily exploited in a higher temperature working range by simply replacing the Al with higher critical temperature superconductors, such as Nb\cite{Angers2008a} or V\cite{Ronzani2013,Ronzani2014}. Hence, the fruition of this identical technology to regimes of operation ranging between 4K to 10K is achievable Finally, our devices exhibit a very promising  flux-to-voltage response characteristics which, in this prototypical realization, let us to foresee a performance already  on pair with that of bi-SQUIDs arrays composed of hundreds of tunnel junction-based interferometers.

\begin{figure}[t]
  \includegraphics[width=\columnwidth]{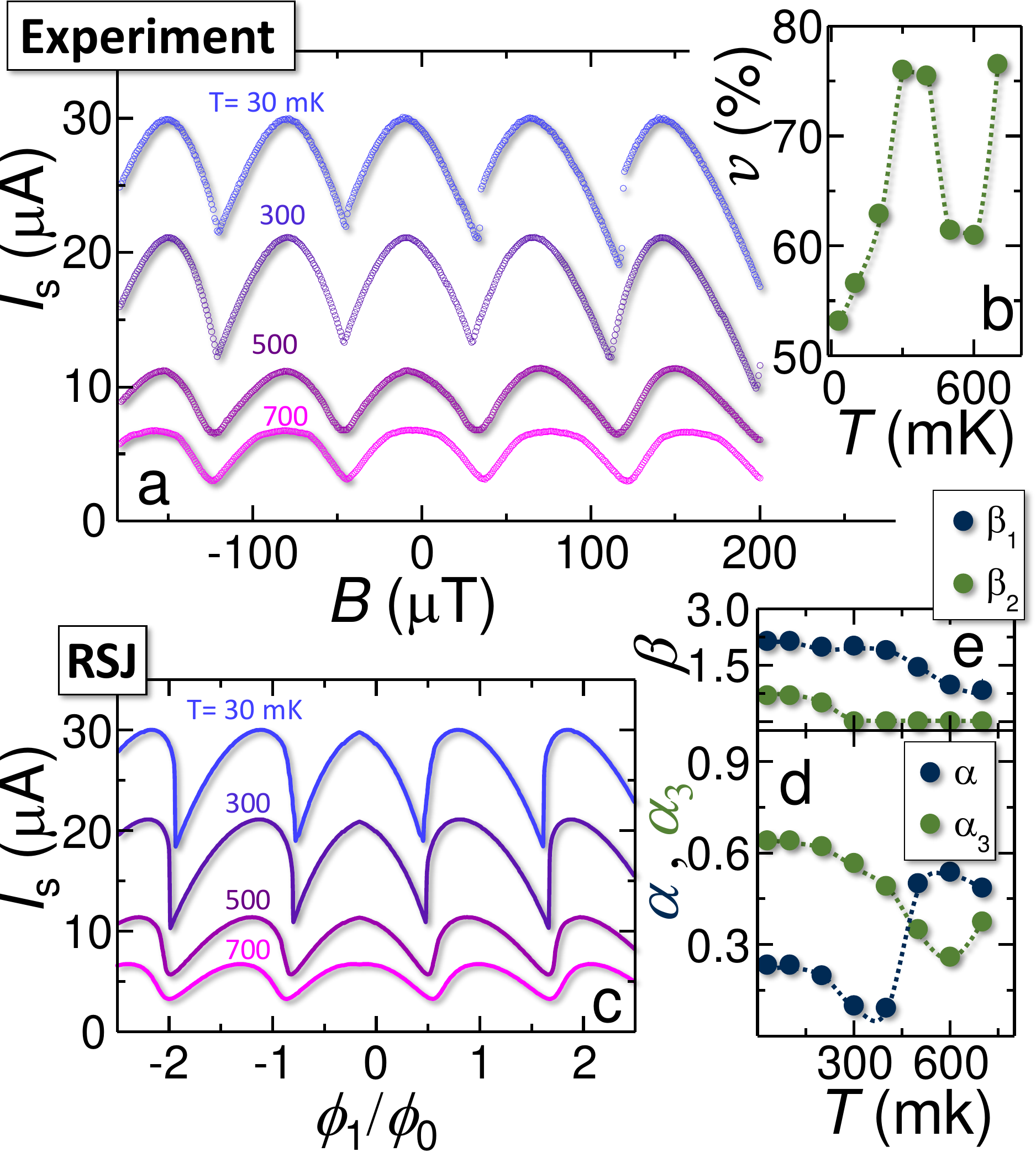}
    \caption{\textbf{Switching current \textit{vs.} flux characterization of the SNS SQUID.} 
    \textbf{a}: Switching current $I_S$ of the bi-SQUID as a function of the out of plane magnetic field  $B$ for selected bath temperatures ranging from 30 mK to 700 mK. The external out-of-plane magnetic field was applied through a superconducting electromagnet. $I_S(B)$ is shown for selected temperatures ranging between 30 mK and 700 mK. 
    \textbf{b}:  Critical current modulation visibility $v=2(I_{S_{max}}-I_{S_{min}})/(I_{S{max}}+I_{S_{min})}$ \textit{vs.}  temperature $T$. 
    \textbf{c}: Plot of the the numerical $I_S(\phi)$ obtained through a RSJ fit of the experimental data of Fig. \ref{fig:fig2}a. 
    \textbf{d}:  Asymmetry parameters ($\alpha$, blue dots) and ($\alpha_3$, green dots) as a function of  the temperature.
    \textbf{e}:  Screening parameters ($\beta_1$, blue dots) and ($\beta_2$, green dots) as a function of  temperature. The values reported in panel d and e are extracted through the fitting procedure (see text).
    } 
  \label{fig:fig2}
\end{figure}

\section{Results and Discussion}
A false-colored scanning  electron micrograph of a representative SNS bi-SQUID is shown in Fig. \ref{fig:fig1}a: it consists of a main 100-nm-thick Al (colored in blue) superconducting loop (loop $1$), spanning an area of $\sim$30 $\mu$m$^2$, closed on $\sim$10 times smaller loop (loop $2$) and comprising  three 20-nm-thick and 320-nm-long Cu weak-links (brown colored in Fig \ref{fig:fig1}b) named A, B and C, according to the labelling reported in Fig. \ref{fig:fig1}b. Junctions A and B were $\sim120-$ nm-wide, while junction C was 90 nm-wide. All the junctions were meant to fall in the diffusive regime and within the long-junction limit. Indeed, the Thouless energy of the junctions $E_{Th}$ is  $E_{Th}=\frac{\hbar D}{l^2}\simeq 50$ $\mu$eV$ \ll \Delta_{Al} \simeq 180$ $\mu$eV, where $D\simeq0.008$ m$^2$/s is the Cu diffusion coefficient \cite{DeSimoni2019}, $l$ the weak-link length, and $\Delta_{Al}$ the superconducting energy gap of the Al banks. Further  details  of the fabrication  process  are reported in  the \textit{Methods} section. 

Figure  \ref{fig:fig1}c  shows  the $IV$ characteristics of a bi-SQUID measured as function of the bath temperature, ranging  from  30  mK  to  900  mK.  A scheme of the 4-wire electrical setup is displayed in  Fig. \ref{fig:fig1}a. For temperatures lower than $\sim 800$ mK, the $IV$s  shows the Josephson effect with the switching $I_S$ and the retrapping current\cite{Courtois2008,Dubos2001} $I_R$ reaching $\sim 30$ $\mu$A and $\sim7$ $\mu$A, respectively, at 30 mK. At the same temperature the normal-state resistance $R_N$ is $\sim 40\, \Omega$. $I_S$ and $I_R$ are plotted \textit{versus}  bath temperature ($T$) in Fig. \ref{fig:fig1}d. 

The measurement of the $IV$ curves as a function of an external out-of-plane magnetic field $B$ allows to reconstruct the $I_S(\phi_1, \phi_2)$ characteristics of the  bi-SQUID, where $\phi_1$ and $\phi_2$ are the magnetic fluxes threading the loop 1 and 2 respectively (see Fig. \ref{fig:fig1}a). The $I_S(B)$ of a representative device is reported in Fig. \ref{fig:fig2}a for selected temperatures ranging between 30 mK and 700 mK. Critical current modulation are present up to $\sim 850$ mK, with a modulation visibility. $v=2(I_{S_{max}}-I_{S_{min}})/(I_{S{max}}+I_{S_{min})}$ ranging from $\sim52\%$ (at 30 mK) to $\sim75\%$ at (700 mK). $v$ is plotted \textit{versus} the temperature $T$ in Fig.~\ref{fig:fig2}b. This confirms once more the good sensing performances of these devices.

Similarly as for single-loop SNS interferometers, due to the contribution to the total flux of the second loop and also to the presence of a sufficiently small junction capacitance (C), bi-SQUID current-flux relation can still be described within the framework of the resistively-shunted junction (RSJ) model\cite{Clarke2004}. However, the behaviours observed for bi-SQUIDs deviate significantly from the ones observed in conventional DC SQUIDs. Indeed, the RSJ model, although conceived for Josephson tunnel junctions, retains its validity for weak-links with a sinusoidal current-phase relation like~\textit{long} SNS junctions and can, therefore, be adapted to our SNS bi-SQUID circuit by merely including the equations accounting for magnetic flux quantization on both the loops as follows: 
\begin{equation}
    i=I_0 [(1+\alpha) \sin(\delta_A)+(1-\alpha) \sin(\delta_B)],
\end{equation}
\begin{equation}
    \delta_C=\pi \beta_1 j_1 - 2\pi \phi_1 /\phi_0=\pi \beta_1 j_2 -2\pi \beta_1 \alpha_3 \sin(\delta_c)- 2\pi \phi_1 /\phi_0,
\end{equation}
\begin{equation}
    \delta_C=\delta_B - \delta_A + 2 \pi \phi_2 / \phi_0 -\pi \beta_2 j_2,
\end{equation}
where $\phi_0\simeq 2.067 \times 10^{-15}$ Wb is the magnetic flux quantum,  $\alpha=\frac{I_A-I_B}{I_A+I_B}$ describes the asymmetry between the critical currents $I_A$ and $I_B$ of the two Josephson junctions A and B, $I_0=\frac{I_A+I_B}{2}$ corresponds to one half the critical current of the interferometer (\textit{i.e.} the average critical current of junctions A and B), and $\alpha_3=\frac{I_C}{I_0}$ accounts for the critical current $I_C$ of the third junction. $\delta_{A,B,C}$ are the phase differences across the weak links, $i$ and $j_{1,2}$ are the supercurrent passing-through the bi-SQUID, and the current circulating in the loop 1 and 2, respectively. Finally, through the screening parameter $\beta_{1,2}$ the rings inductances are accounted for\cite{Clarke2004}. By means of a fit of the $I_S(B)$\cite{Ronzani2013,DeSimoni2021}  curves against the RSJ model, we extracted the relevant device parameters at 30 mK such as the effective loop areas ($\sim22 \pm 0.7$ $\mu$m$^2$ and  $\sim2 \pm 0.7$ $\mu$m$^2$), the asymmetry parameters ($\alpha=0.23 \pm 0.1$ and $\alpha_3=0.6 \pm 0.2$), and the screening parameters ($\beta_1=2.1 \pm 0.7$ and $\beta_2=0.7 \pm 0.7)$. Such values are in agreement with the design of our SNS bi-SQUID. Their evolution with temperature is shown in panels d and e of Fig. \ref{fig:fig2}. Both $\beta_1$ and $\beta_2$ decays with the temperature due to the decay of the critical current of the junctions. In particular $\beta_2$ becomes negligible at $\sim300$ mK, due to the faster decay of $I_C$ with respect to $I_A$ and $I_B$. This fact is confirmed by the behavior of $\alpha_3$ \textit{versus} $T$, which decreases from the value of $\sim0.6$ to about 0.3 at 600 mK. The $I_S(\phi_1)$ curves calculated from the fitting procedure are shown in Fig. \ref{fig:fig2}c (note that $\phi_2$ is just proportional to $\phi_1$). All of this confirms the ability of the RSJ model to successfully capture the essential physical aspects that are into play in these devices.  

We now discuss the performance of our SNS bi-SQUIDs in view of its possible exploitation as a linear-response magnetic flux sensor operating in the dissipative regime, which is realized by current biasing the interferometer across or above its switching current at a fixed magnetic field working point.  The modulation of the magnetic flux is transduced into  a  change of the voltage drop $V$ developed at the ends of the interferometer.

\begin{figure*}
  \includegraphics[width=\textwidth]{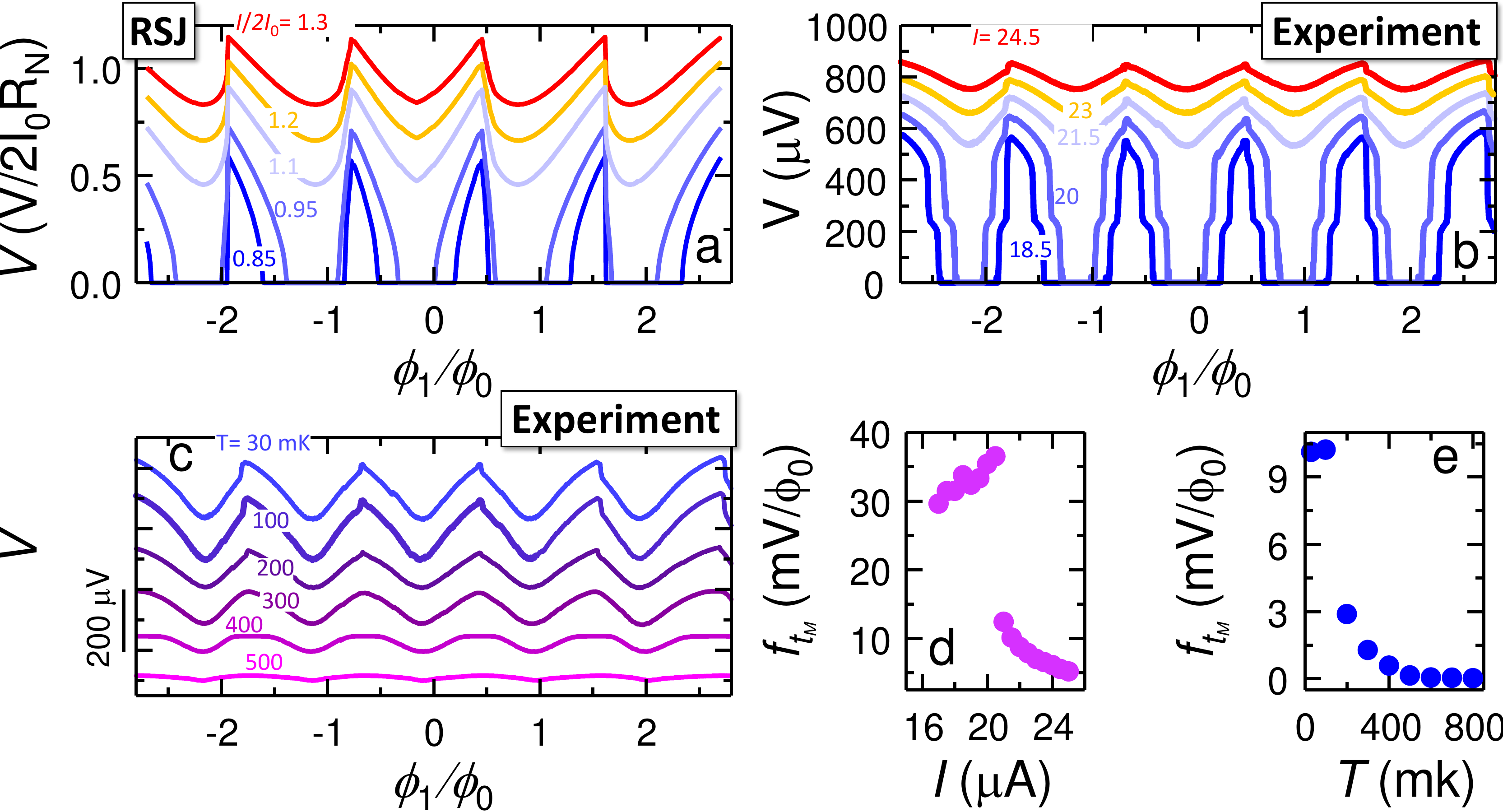}
    \caption{
    \textbf{SNS bi-SQUID operated in the dissipative regime.}
    \textbf{a}: Voltage-flux response curves numerically calculated through an RSJ model ($V(\phi_1)=R_{N}\sqrt{I^2-I_S^2}$) at selected relative current bias $I/I_S$. The curves were calculated by means of the numerical $I_S(\phi_1)$ derived from the fit of the experimental data of Fig. \ref{fig:fig2}a at 30 mK.
    \textbf{b}: 4-wire lock-in $V(\phi_1)$ characteristics at 30 mK for  selected values of the room-mean-square bias current $I$ between 3 $\mu$A and 6.4 $\mu$A, measured by biasing the device with a 17 Hz sinusoidal current signal. Below $I\simeq 21.5$ $\mu$A, the curves show null voltage-drop when $I<I_S(\phi_1)$. A finite $V$ value is measured when the device is in the dissipative regime, \textit{i.e.} when $I$ is larger than the flux-dependent switching current.
    \textbf{c}: 4-wire lock-in $V(\phi)$ characteristics at selected temperature between 30 mK and 500 mK. $I$ was chosen at each temperature in order to let the device to operate permanently in the dissipative regime, and to maximize the linearity of the voltage-flux response (see text). $I$ was set to 21.5 $\mu$A, 21 $\mu$A, 20 $\mu$A, 16 $\mu$A, 13 $\mu$A, and 9 $\mu$A for $T$ respectively equal to 30 mK, 100 mK, 200 mK, 300 mK, 400 mK, and 500 mK.
    \textbf{d}: Maximum value $f_{t_M}$ of the transfer function ($f_t$)  \textit{vs} $I_{RMS}$. 
    \textbf{e}: Maximum value $f_{t_M}$ of the transfer function ($f_t$)  \textit{vs} $T$. 
    }
  \label{fig:fig3}
\end{figure*}
\begin{figure}[ht]
  \includegraphics[width=0.9\columnwidth]{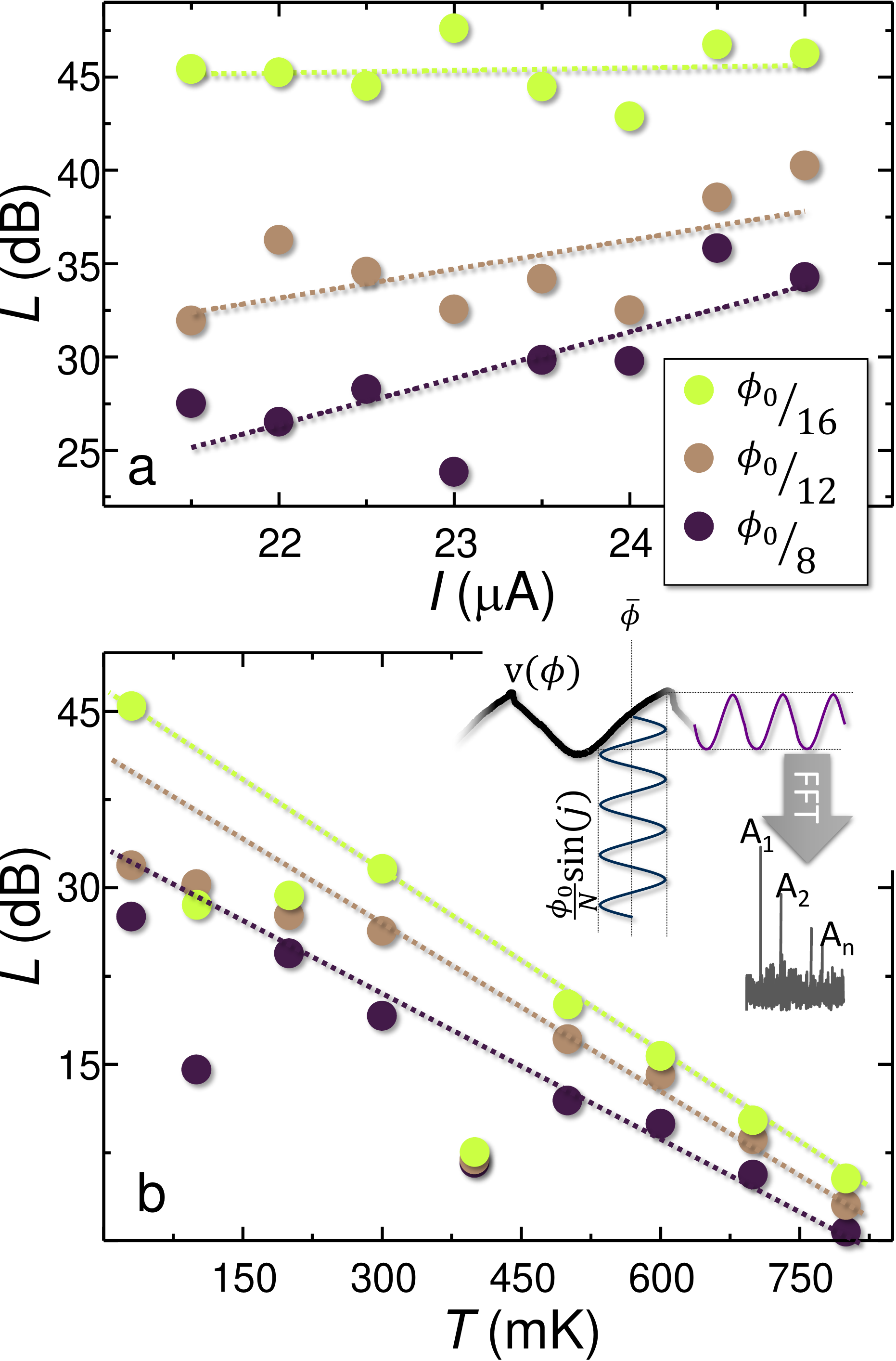}
    \caption{\textbf{Voltage-flux response linearity of the SNS bi-SQUIDs extracted from the experimental data}
    \textbf{a}: Total harmonic distortion $L=-20 log(\frac{\sqrt{\sum_{n=2}^{\infty} A_n^2}}{A_1})$ $vs.\,I$ for selected values of the flux modulation amplitude $\Phi$, between $\phi_0/8$ and $\phi_0/16$. Dashed lines are guides for the eye.
    \textbf{b}: Temperature $T$ evolution of the linearity $L$. \textbf{Inset}: Schematic of the method exploited to extract $L$ from the $V(\phi_1)$ characteristics. A sinusoidal modulation $\Phi sin(\omega_1)$ of the flux around the working point $\overline{\phi}$ with arbitrary frequency $\omega_1/2 \pi$ translates into a voltage signal through the experimental flux-to-voltage response curves (plotted in Fig. \ref{fig:fig3}b and c). The resulting signal is then fast-Fourier transformed to compute $L$.   
    }
  \label{fig:fig4}
\end{figure}

Within the RSJ framework, it is possible to deduct  the $V(\phi)=R_{N}\sqrt{I^2-I_S^2}$  voltage response curve \cite{Clarke2004} at constant current bias $I$, through the knowledge of the device switching-current \textit{vs.} flux characteristics and the device total resistance in the normal state $R_N$. Figure \ref{fig:fig3}a shows the $V(\phi)$s extracted from the numerical curve $I_S(\phi_1)$ at 30 mK (see Fig. \ref{fig:fig2}c) for selected values of the normalized bias current $I/2I_0$. Such theoretical $V(\phi_1)$ curves are significantly different from those of conventional SQUIDs, showing, indeed, a \textit{shark-fin} voltage oscillation in the flux, for current bias above the switching current. At lower bias currents, zero voltage-drop regions are observed for magnetic fluxes and bias current such that $I<I_S(\phi_1)$. This results in a strongly nonlinear behaviour at the switching points, and in quasi-linear regime in the finite-resistance regions. 
The effective linearity ($L$) in the latter can be quantified by the figure of merit of the total harmonic distortion (THD)\cite{Soloviev2019}, which is here defined, through the Fourier transform of the device voltage response to a sinusoidal modulation of the magnetic flux around a fixed working point $\overline{\phi}$ such that $\phi_1=\overline{\phi}+\Phi sin(\omega_1)$, as $L=-20 log(\frac{\sqrt{\sum_{n=2}^{\infty} A_n^2}}{A_1})$, where $A_n$ is the amplitude of the $n^{th}$ harmonic of the excitation signal, and $\Phi$ is the amplitude of the flux modulation (see the inset of Fig. \ref{fig:fig4}b). Such method, which is valid in the $\omega_1 \rightarrow 0$ limit, yields a theoretical maximum value for $L$ ranging from $\sim70$ dB (at $I_B/I_0=0.85$) to $\sim120$ dB (at $I_B/I_0=1.3$), for $\Phi=\phi_0 /16$. As discussed in many other previous works~\cite{Kornev2009,Kornev2011,Kornev2011a,Kornev2014,Kornev2017,Kornev2017a,Kornev2020,Kornev2020a}, the inability associated to  many real bi-SQUIDs technologies to achieve performances similar to the ones anticipated theoretically is a critical problem. One of the main results in this paper is that the technology we propose is one of the first that enable real performance for a single bi-SQUID close to the theoretical ones. 

The characterization of a representative bi-SQUID in the dissipative regime is reported in Fig. \ref{fig:fig3}b where we plot the voltage drop across the device at 30 mK recorded by a 4-wire lock-in technique for selected amplitudes of a 17 Hz sinusoidal current-bias signal. 
Similarly to the RSJ model, below  $I\sim 21.5\mu$A, the curves exhibit  a zero voltage-drop for magnetic fluxes such that $I<I_S(\phi_1)$, while a finite $V$ value is measured when the interferometer is driven into the dissipative regime. For $I\gtrsim 21.5$ $\mu$A the device permanently operates in the latter, showing the same \textit{shark-fin} behavior as observed theoretically with the RSJ model. This regime can be fully exploited up to above $\sim500$ mK, with a voltage-swing amplitude increasing from $\sim15$ $\mu$V (at 500 mK) to $\sim200$ $\mu$V by lowering the temperature down to 30 mK (see Fig. \ref{fig:fig3}c). For temperature lower than $\sim300$ mK the shape of the $V(\phi_1)$ characteristics retains the \textit{shark-fin} shape observed at 30 mK. At higher temperatures such a behavior evolves into a \textit{trapezoidal}  oscillation of the voltage \textit{vs.} flux characteristics.

The conventional figure of merit of the maximum value of the transfer function $f_{t_M}$, calculated through numerical differentiation of the $V(\phi)$, \textit{vs.}  $I$ and $T$, is also shown in Fig. \ref{fig:fig3}d and e, respectively. $f_{t_M}$ linearly grows as a function of the bias current up to a value of $\sim35$ mV/$\phi_0$ at $\sim20$ $\mu$A. Above such a value, the device is fully resistive, resulting in a sudden drop of the transfer function and in a fast decay of $f_{t_M}$ with the biasing current. Finally, $f_{t_M}$ decays with the temperature from the value of 10 mV/$\phi_0$ obtained at 30 mK, and vanishes for $T \gtrsim 500$ mK. On this point we emphasize that, in terms of the transfer function, our SNS bi-SQUID outperforms by two orders of magnitude interferometers of similar typology \cite{Ronzani2013,DeSimoni2021}.

Finally, We focus on the measure of the linearity of the low-frequency voltage-flux response of SNS bi-SQUIDs. This was extracted from the experimental data by following the same numerical approach exploited to assess the THD of a sinusoidal modulation of the magnetic flux by using the $V(\phi_1)$ RSJ characteristics and schematically depicted in the inset of Fig. \ref{fig:fig4}b. Figure \ref{fig:fig4}a reports $L\, vs.\, I$ for selected values of the flux modulation amplitude $\Phi$, between $\phi_0/8$ and $\phi_0/16$. $L$ exhibits a linear dependence on $I$ and ranging between 25 dB at 21 $\mu$A (and $\Phi=\phi_0/8$), and $\sim46$ dB at 25 $\mu$A (and $\Phi=\phi_0/16$). By decreasing the amplitude of $\Phi$, the average value of $L$ was observed to increase due to the reduction of the ratio between the linear portion of the $V(\phi_1)$ characteristics and $\Phi$. Furthermore the slope of $L\, vs.\, I$  decreases by reducing the flux modulation amplitude. This suggests that, for practical applications, at higher values of $\Phi$ the bias current might be exploited as knob to increase the linearity of the device response. $L$ persists within the range between 20 dB and 45 dB up to $\sim 300$ mK. Above such threshold temperature it linearly decreases as a consequence of the temperature evolution of the $V(\phi_1$) which moves from the \textit{shark-fin} shape to the \textit{trapezoidal} one. 
We conclude this section by remarking the relevance of the performance achieved by our SNS bi-SQUID  in terms of response linearity. Indeed, the measured $L$ values, although still far from those expected from the RSJ model, yet are very promising when compared to those of tunnel junction-based devices composed of arrays of hundreds of interferometers\cite{Kornev2020a}.

\section{Conclusions}
Among available magnetic field sensors, superconducting interferometers have progressively become an essential tool for a multitude of applications ranging from  nanomagnetometry to  telecom signal processing. Here, we have discussed the implementation of a superconducting bi-SQUID based on proximitized mesoscopic Al/Cu/Al junctions. Such a design provides a viable technological alternative to the conventional approach based on tunnel JJs, and promises an excellent performance in terms of transfer function as well as of voltage-to-flux response. The latter, which we measured in the low-frequency limit, is on par with the performance obtained so far  on tunnel junction-based devices made of several tenths or hundreds of bi-SQUIDs. This consideration let us to speculate about a successful exploitation of SNS bi-SQUIDs in place of tunnel junction-based interferometers. 

%%%%%%%%%%%%%%%%%%%%%%%%%%%%%%%%%%%%%%%%%%%%%%%%%%%%

\section*{Methods}
\subsection{Device nanofabrication}
The SNS bi-SQUIDs were fabricated in a single electron-beam lithography (EBL) step and a two-angle shadow-mask metal deposition through a suspended PMMA resist mask onto a SiO$_2$ substrate. The Al/Cu SN clean interfaces were obtained through electron-beam evaporation in an ultra-high vacuum (UHV) chamber with a base pressure $\sim5\times 10^{-11}$ Torr. 
The Ti/Cu bilayer (with thickness 5/25 nm, and Ti promotes adhesion) was evaporated at an angle of $0^\circ$ to realize the SQUID nanowires. The sample holder was then tilted at an angle of $13^\circ$ for the deposition of a 100-nm-thick film of Al to realize the superconducting loop.

\subsection{Cryogenic electrical characterization}
The  electrical  characterization of the interferometer  was  carried out  by  four-wire technique  in a  filtered cryogen-free $^3$He-$^4$He dilution  fridge equipped with a superconducting electromagnet. Current-voltage ($IV$) measurement were carried out by setting a low-noise current bias through a room temperature voltage generator and a bias resistor. The voltage drop across the interferometer was measured with a room temperature pre-amplifier. Switching current values were averaged over the switching points of 15 repetitions of the same $IV$. The voltage $vs.$ flux characterization was performed through a low-frequency lock-in technique.

%\input{text_1st_revised}
%\begin{acknowledgement}
\section*{Acknowledgement}
The authors  acknowledge the European Research Council under Grant Agreement No. 899315-TERASEC, and  the  EU’s  Horizon 2020 research and innovation program under Grant Agreement No. 800923 (SUPERTED) and No. 964398 (SUPERGATE) for partial financial support.  F.G. and G.C.T. acknowledge many useful conversations on the matters of this research with Dr P. Atanackovic.

%\end{acknowledgement}

\section*{Author Contributions}
N.L. fabricated the devices. L.C. and G.D.S performed the experiment with input from F.G.. L.C and G.D.S. analyzed the data with inputs from F.G.. G.D.S implemented the R.S.J. numerical model with inputs from all the authors. G.D.S. wrote the manuscript with input from all the authors. F.G. and G.T. conceived the experiment. All of the authors discussed the results and their implications equally.

%\bibliography{Bibliography}

\end{document}